\documentclass[3p,times,procedia]{elsarticle}

\usepackage{ecrc}
\usepackage{algorithm}
\usepackage{algorithmic}
\usepackage{subfigure}

\volume{00}

\firstpage{1}

\journalname{Procedia Computer Science}

\runauth{}

\jid{procs}

\jnltitlelogo{Procedia Computer Science}

\CopyrightLine{2011}{Published by Elsevier Ltd.}

\usepackage{amssymb}

\usepackage[figuresright]{rotating}

\begin{document}

\begin{frontmatter}

\dochead{International Workshop on Body Area Sensor Networks (BASNet-2013)}

\title{Energy Consumption Rate based\\ Stable Election Protocol (ECRSEP) for WSNs}

\author{O. Rehman$^{\pounds}$, N. Javaid$^{\pounds}$, B. Manzoor$^{\pounds}$, A. Hafeez$^{\pounds}$, A. Iqbal$^{\pounds}$, M. Ishfaq$^{\S}$}

\address{$^{\pounds}$COMSATS Institute of Information Technology, Islamabad, Pakistan. \\
        $^{\S}$King Abdulaziz University, Rabigh, Saudi Arabia.}

\address{}

\begin{abstract}
In recent few years Wireless Sensor Networks (WSNs) have seen an increased interest in various applications like border field security, disaster management and medical applications. So large number of sensor nodes are deployed for such applications, which can work autonomously. Due to small power batteries in WSNs, efficient utilization of battery power is an important factor. Clustering is an efficient technique to extend life time of sensor networks by reducing the energy consumption. In this paper, we propose a new protocol; Energy Consumption Rate based Stable Election Protocol (ECRSEP). Our CH selection scheme is based on the weighted election probabilities of each node according to the Energy Consumption Rate (ECR) of each node. We compare results of our proposed protocol with Low Energy Adaptive Clustering Hierarchy (LEACH), Distributed Energy Efficient Clustering (DEEC), Stable Election Protocol (SEP), and Enhanced SEP(ESEP). Our simulation results show that our proposed protocol, ECRSEP outperforms all these protocols in terms of network stability and network lifetime.
\end{abstract}


\end{frontmatter}


\section{Background and Motivation}
\label{}
%

Clustered sensor networks can be classified into two broad types; homogeneous and heterogeneous sensor networks. In homogeneous networks all sensor nodes are identical in terms of energy and hardware complexity. With purely static clustering in a homogeneous network, it is evident that CHs will be over-loaded with long range transmissions to the remote sink, and extra processing is necessary for protocol co-ordination and data aggregation. WSN faces a problem that CHs dies before other nodes. However, to ensure that all nodes dies at about the same time when system expires, minor amount of residual energy is wasted. One method to ensure is rotating the role of a cluster head periodically and randomly over all the nodes. The downside of role rotation and using a homogeneous network is that all nodes should be capable of act as CH, therefore should require necessary hardware capabilities. On the other hand, in heterogeneous sensor network, two or more different types of sensor nodes in terms of different energy are used. The problem area is that extra energy and complex hardware can be embedded in few CH nodes, therefore reducing hardware cost of the entire sensor network.


In LEACH the sensor nodes are equipped with same amount of energy. This protocol selects CH periodically and consumes energy uniformaly. Each node is decide itself whether or not a CH based on probability [1]. SEP is based on two level heterogeneity and the CH election in SEP is on the basis on weighted election probability. A fraction $ m $ advanced nodes in total of $ n $ nodes is provided with an additional energy factor $ \alpha $. So, the stability period is increased due to advance nodes, however CH selection is done in the same way as in LEACH. ESEP is an extension of SEP that considers three types of nodes in discussed in [2,3]. In [4] DEEC estimates ideal value of network life time is used to compute reference energy that each node expend during a round.


\begin{figure}
\begin{center}
\includegraphics[height=5cm,width=10cm,angle=0]{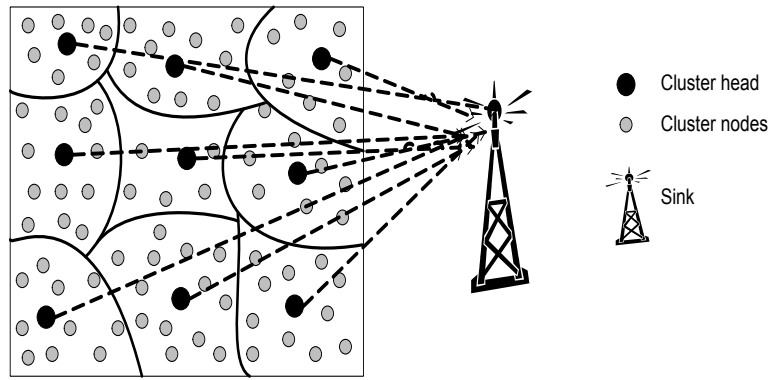}
\caption{\small \sl Cluster Formation in WSN.\label{fig:Stupendous}}
\end{center}
\end{figure}

In WSN, current scenario of research deals with efficient power utilization of sensor nodes. Due to small battery power of these nodes, there is a chance that WSN is not longer survive. Smart utilization of sensor node is very crucial to prolong the life time and stability of WSN. Most current protocols, such as SEP and ESEP are stability-oriented protocols, minimize energy utilization of network by using clustering approach. stability period is the time interval before the death of first node in WSN. In clustering data is transmitted to sink in form of clusters and every cluster consists of a CH, responsible to transmit data at sink.

Existing protocol SEP describe impact of heterogeneity on heterogeneous-aware protocols and instability of protocols, such as LEACH in presence of heterogeneity of sensor nodes. SEP is based on weighted election probabilities assigned to each node to become CH according to their initial energy. The rotating epoch and election probability is directly correlated with initial energy of nodes instead of residual energy of the nodes. Advance nodes more frequently becomes CHs, it may happen after some rounds the energy of advance nodes becomes less than normal nodes. To overcome this drawback, we introduce a new CH selection scheme for SEP based on ECR of each node. By using this criteria SEP increases stability and lifetime of network.

\section{SEP}
SEP improves the stable region of a WSN by using the heterogeneity  parameters such as fraction of advanced nodes $ m$ and additional energy factor $\alpha $ between the normal and advance nodes. To prolong the stability region of a network, SEP maintain the constraints of well balance energy consumption.\\
In SEP initially, advanced nodes have to become the CH more often than normal nodes.
suppose that $ E_0$ is the initial energy of each normal node and $ E_0(1+\alpha)$ is the energy of advanced nodes in a WSN. The total energy of new heterogeneous network in [2] is equal to: $n.(1-m).E_0 + n.m.E_0 (1+\alpha)=n.E_0.(1+ \alpha .m)$. Total energy of the system in increased by $ 1+\alpha.m $ times. In order to increase the stability of the system, new epoch must equal to $ \frac{1}{p_{opt}}(1+ \alpha.m)$ because system has $ \alpha. m $ times more nodes and $ \alpha.m $ more energy.
Initially, for each node the probability of becoming CH is $ p_{opt} $ . An average $ n \times p_{opt}$ must becomes CHs per round per epoch. The nodes that are elected to be CH in current round can no longer become CH in the same epoch.Nodes that are not elected CHs belongs to set $ G $ in order to maintain a steady number of CHs per round. The probability of nodes $ s \epsilon  G$ to become CH is increases after each round in same epoch. The decesion is made at the beginning of each round by each node $ s \epsilon  G$ independently chossing a random number between [ 0,1 ]. If random number is less than threshold $ T(s) $ then the node become a cluster head in current round.
The threshold is set in [2] as:
\begin{eqnarray}
T_{S} = \left\{ \begin{array}{rl}
 \frac{p_{opt}}{1-p_{opt}[r. mod \frac{1}{p_{opt}}]} &\mbox{ if $ s \epsilon  G$} \\
  0 &\mbox{ otherwise}
       \end{array} \right.
\end{eqnarray}
where r is the current round number.

 SEP increase the stable region of a network, if fulfilling the following conditions.\\
 a. Each normal nodes becomes a CH once every $\frac {1}{p_{opt}}.(1 + \alpha.m)$ rounds per epoch.\\
 b. Each advanced node becomes a CH $ 1 + \alpha $ times every $\frac {1}{p_{opt}}.(1 + \alpha.m)$ rounds per epoch. \\
 c. Average number of CH per round per epoch is equal to $ n \times p_{opt}$.\\
   If at the end of each epoch the number of times that an advanced node becomes CH is exactly not equal to $ 1+\alpha$ times, so energy is not well distributed and average numbers of CH per epoch per round is not equal to $ n \times p_{opt}$.
 \begin{equation}
p_{adv}=\frac{p_{opt}}{1+\alpha m}
\end{equation}
\begin{equation}
p_{nrm}=\frac{p_{opt}(1+\alpha)}{1+\alpha m}
\end{equation}
\section{ECRSEP}
In ECRSEP, CH selection is based on the Energy Consumptio Rate (ECR).
 ECR is defined mathematically as:
$ ECR = \frac {E_{int}-E_r}{r-1}$.\\
where, $E_{int}$ is initial energy and $E_r$, is residual energy of each node and $r$ is current round.
 In next round, CH selection is based on ECR in previous round. A node, which have less ECR in the previous round  is selected CH in next round. A CH in the previous round is not selected as CH in the next round, because its ECR is very high as compare to non CH nodes.

\subsection{Radio Model}
In radio model, energy dissipates $ E_{elec} = 50nJ/bit $ to run receiver and transmitter circuitry and $ E_{amp} = 100pJ/bit/m^2 $ for transmitter amplifier.
The equations that is used to calculate the receiving cost and transmitting cost for $ k $ bit message and distance $ d $ is modeled in [4] is as shown in below:\\
Transmitter Energy
\begin{equation}
E_{T}(K,d)= E_{T-elect}(k)+E_{T-amp}(k,d)
\end{equation}
\begin{equation}
E_{T}(K,d)=(E_{elect}\times K) + (E_{amp}\times k \times d^2)
\end{equation}

Receiving Energy
\begin{equation}
E_R(K)=E{R-elec}(K)
\end{equation}
\begin{equation}
E_{R}(K)=E_{elec}\times K
\end{equation}
\subsection{Network Model}
In this section, we discuss network model for ECRSEP. Assume that $ N $ sensor nodes are deployed within a $ M \times M $ . The network is deployed into clustering hierarchy. Every cluster has a CH, responsible to directly transmit data to Sink. We suppose that our network is stationary.\\
In our network we considered two level of heterogeneity in terms of energy. In heterogeneous networks, there are two types of sensor nodes, i.e., normal nodes and advance nodes. $ E_O $ is initial energy of normal nodes and $ m $ is fraction of advanced nodes. Advanced nodes have $\alpha$ times more energy than normal nodes. So $mN$ advanced nodes having initial energy $ E_o(1+\alpha)$ and $(1-m)N$ normal nodes having initial energy $E_o$.

The initial energy $ E_o$ of two levels heterogeneous network is given in [2] as:
\begin{equation}
E_{total}=N(1-m)E_o+NmE_o(1+\alpha)=NE_o(1+\alpha m)
\end{equation}
So, two level heterogeneous network have $ am $ times more energy than homogeneous network.

\subsection{CH selection in ECRSEP Protocol}
In this section, we describe the CH selection method in ECRSEP protocol. In this protocol CH selection is based on energy consumption rate.
Let $n$ is the number of rounds to become CH for nodes $S$ that are participating to become CH. we refer to it as rotating epoch.
Let $p=\frac{1}{n}$ is average probability to become CH during $n$ rounds. When nodes have same amount of energy at each epoch, choosing the $ p $ to become $ p_{opt}$ ensures that in every round there are $ p_{opt}N$ cluster-heads, we have
\begin{equation}
p=p_{opt} \times {ECR}
\end{equation}
The total number of CHs per epoch is equal to:
\begin{equation}
\sum_{i}^NP_i=Np_{opt}
\end{equation}

In two level heterogeneous networks, $p_{opt}$ is replaced by weighted probabilities for advance and normal nodes as modeled in [4] as:
\begin{equation}
p_{adv}=\frac{p_{opt}\frac{E(i)-E(r)}{r-1}}{1+\alpha m}
\end{equation}
\begin{equation}
p_{nrm}=\frac{p_{opt}(1+\alpha)\frac{E(i)-E(r)}{r-1}}{1+\alpha m}
\end{equation}
Therefore,$p_{(i)}$ is changed into:

\begin{eqnarray}
p_{(i)} = \left\{ \begin{array}{rl}
 \frac{p_{opt}\frac{E(i)-E(r)}{r-1}}{(1+\alpha m)} &\mbox{ if $S(i)$} \, is\, the\, normal\, node \\
  \frac{p_{opt}(1+\alpha)\frac{E(i)-E(r)}{r-1}}{(1+\alpha m)} &\mbox{ if $S(i)$} \, is\, the\, advance\, node \\
       \end{array} \right.
\end{eqnarray}

We can get probability threshold used to elect CH. Thus the threshold is correlated with energy consumption rate of each node directly.\\

\section{Simulations Result}
We evaluate performance of our protocol by using MATLAB. We arrange a WSN with $ N=100$ nodes are distributed randomly in $ 100m\times 100m$ field. We assume in our simulations that sink is at center of sensing region. To compare performance of ECRSEP with other protocols, effect of interference and signal collision is not considered in wireless channel. Our goal is to compare performance of ECRSEP with SEP, ESEP, LEACH, and DEEC protocol on basis of energy dissipation and the longevity of network.

\begin{figure}[!t]
\centering
\subfigure[Dead Nodes, $\alpha = 2$ and $m = 0.2$]{\includegraphics[height=4.5 cm,width=7 cm]{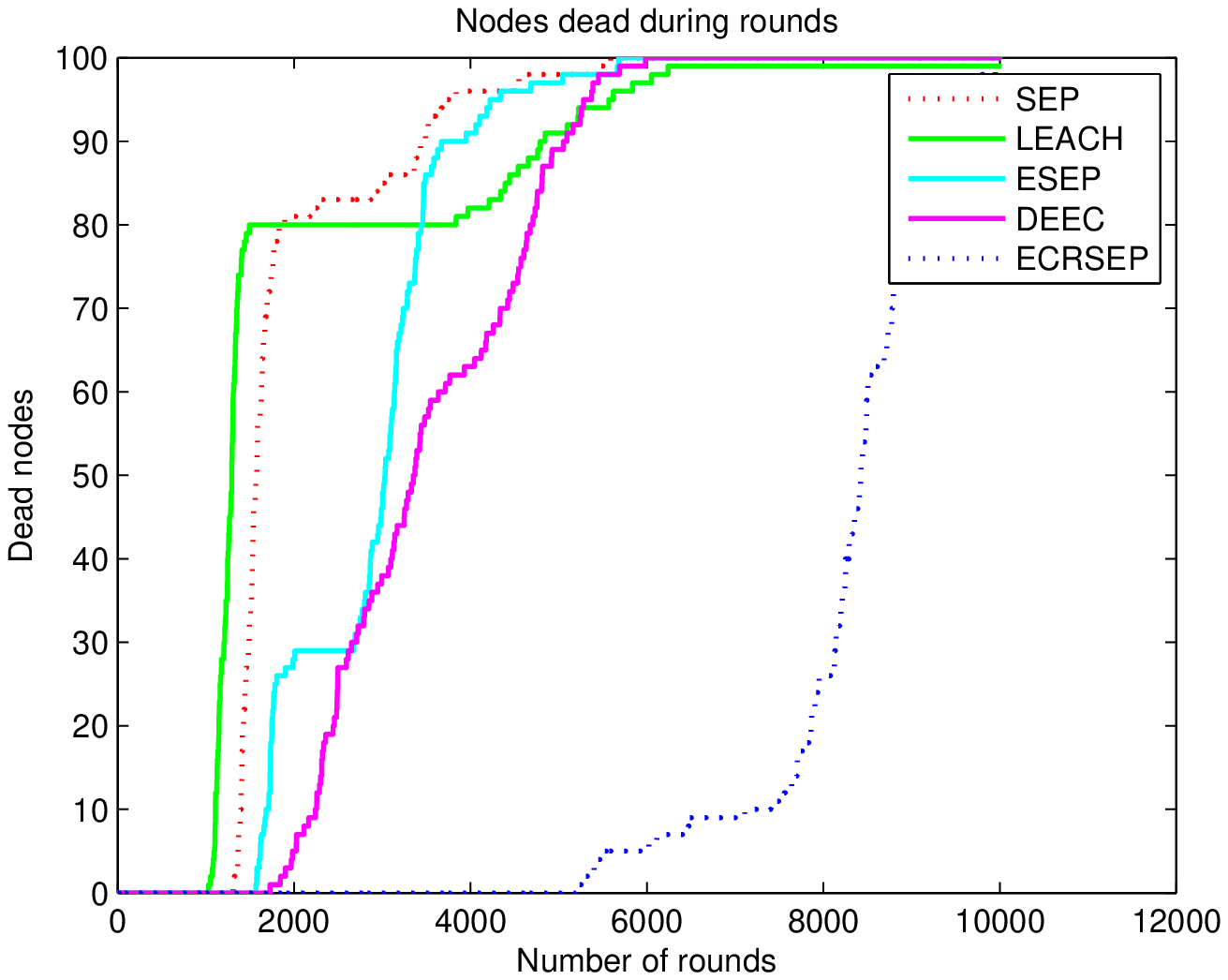}}
\subfigure[Alive Nodes,$\alpha = 2 $ and $m = 0.2$]{\includegraphics[height=4.5 cm,width=7 cm]{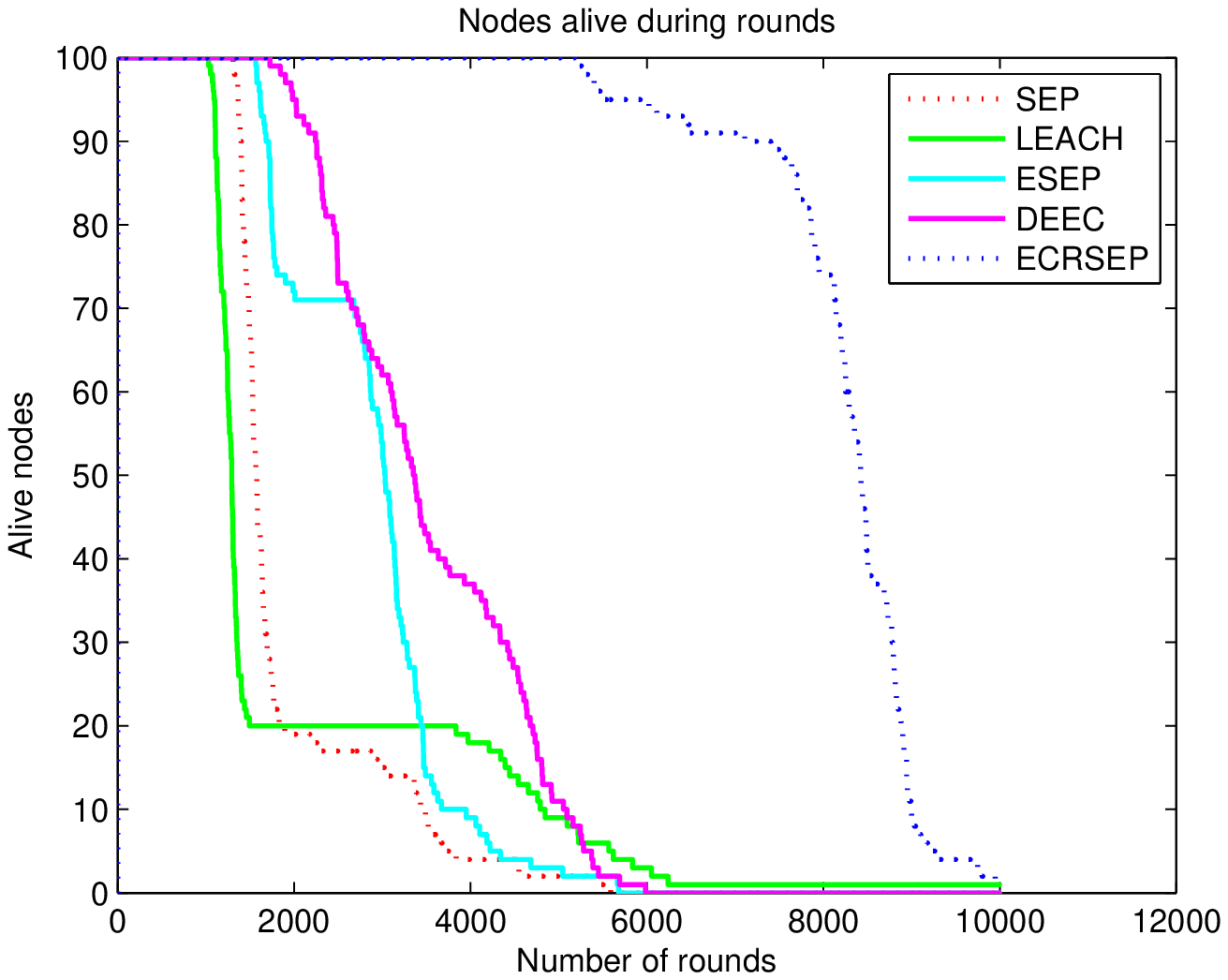}}
\subfigure[Throughput,$\alpha = 2 $ and $m = 0.2$]{\includegraphics[height=4.5 cm,width=7 cm]{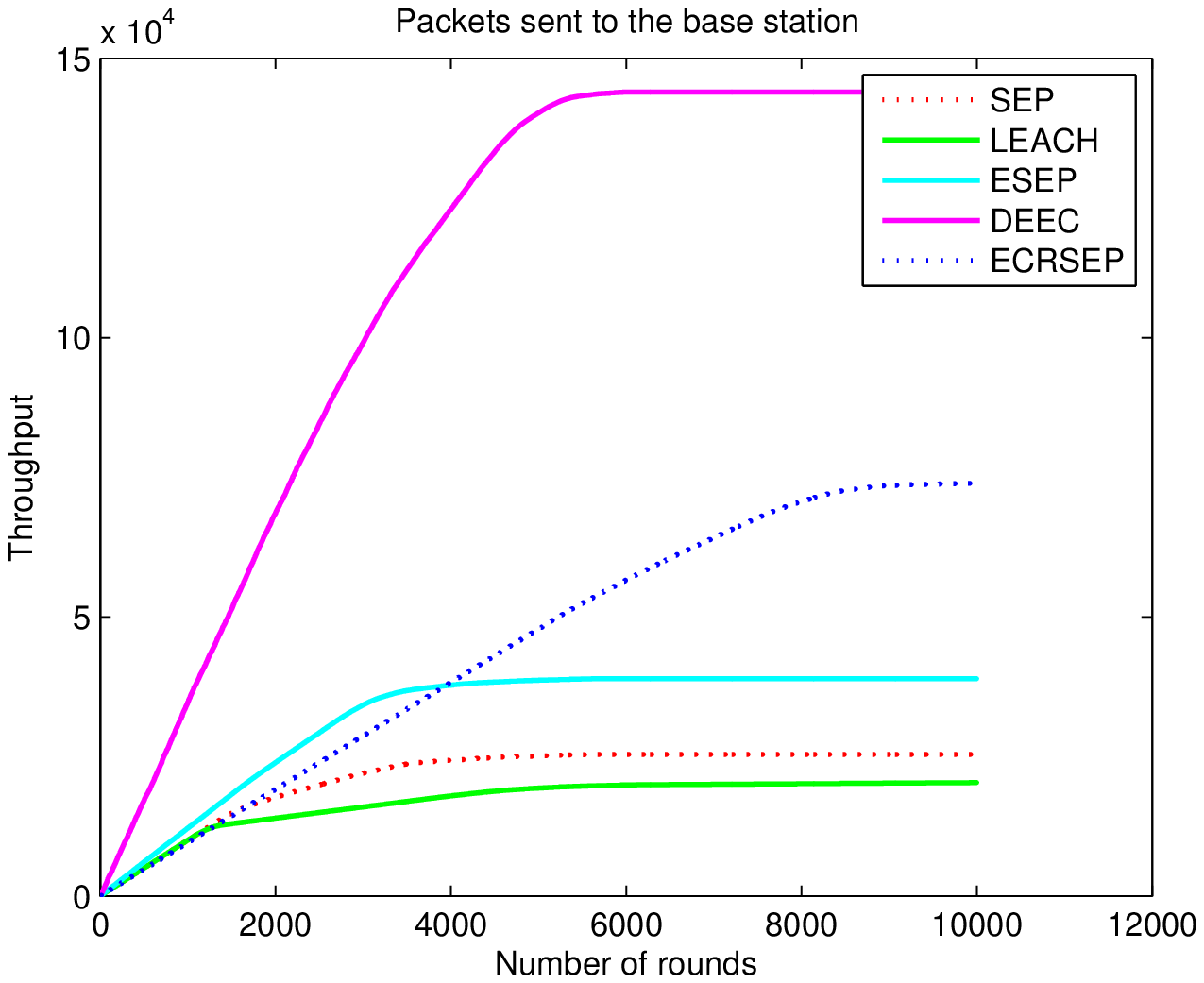}}
\subfigure[Alive Nodes,$\alpha = 2 $ and $m = 0.3$]{\includegraphics[height=4.5 cm,width=7 cm]{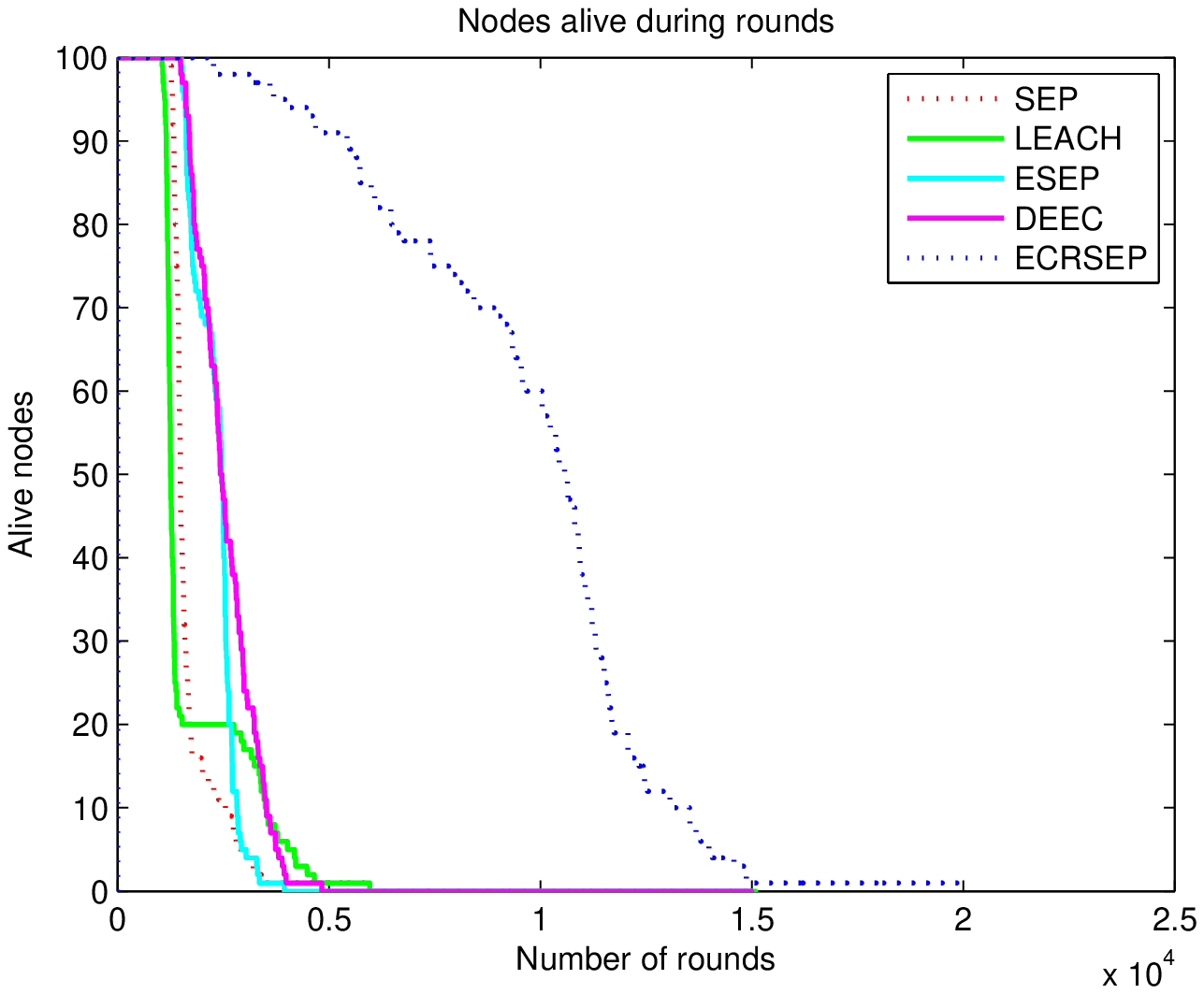}}
\subfigure[Dead Nodes,$\alpha = 2 $ and $m = 0.3$]{\includegraphics[height=4.5 cm,width=7 cm]{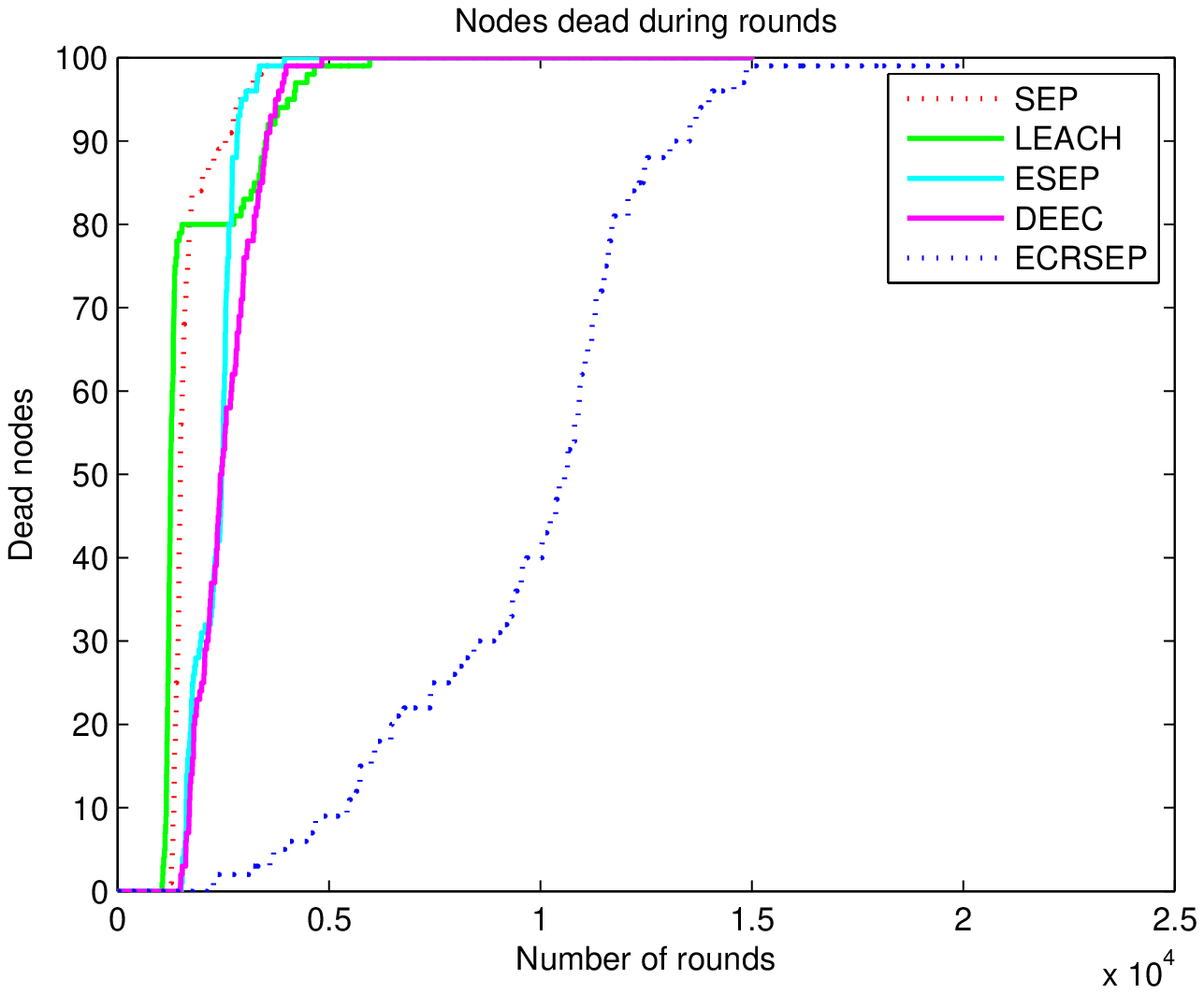}}
\subfigure[Throughput,$\alpha = 2 $ and $m = 0.3$]{\includegraphics[height=4.5 cm,width=7 cm]{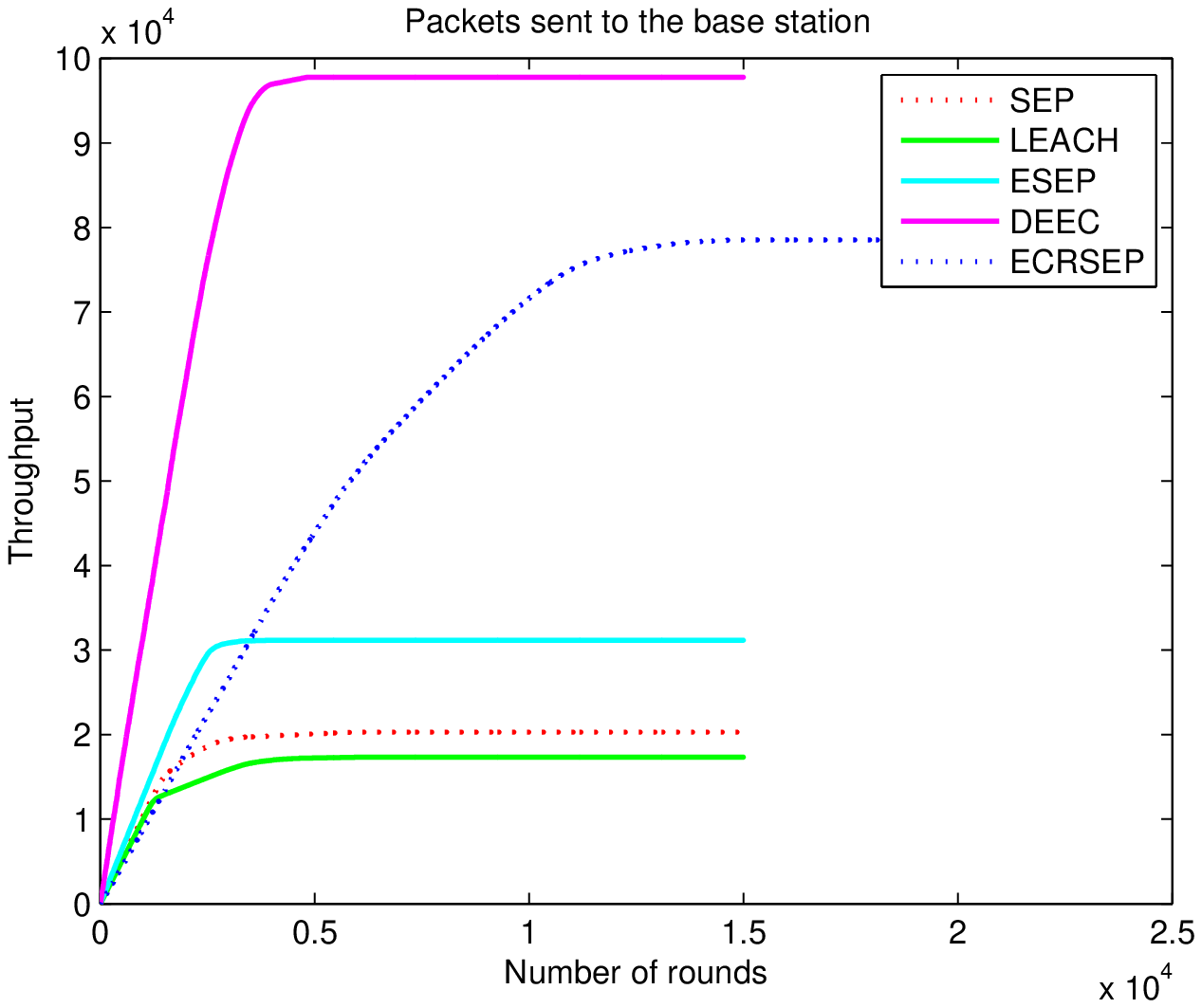}}
\caption{Performance Evaluation of ECRSEP}
\end{figure}
We use following parameters in our simulation. $ E_{elect} =  50nJ/bit$, $E_{DA} = 5nJ/bit/message$,  $\epsilon_{fs} = 10pJ/bit/m^2$ , $\epsilon_{mp} = 0.0013pJ/bit/m^4$, $E_o = 0.5J$, $K = 4000 $, $P_{opt} = 0.1$, $n = 100$, $\alpha = 1$ and $m = 0.1$.

By performing simulations in MATLAB, it is observed that, ECRSEP has enhanced stability period than all other protocol and network life for ECRSEP was increased as compared to others. However DEEC outperforms all protocols in terms of throughput.
The graphs, in Fig. 2 (a,b,c)] results, in a case when $\alpha$ = 2 and $m =0.2$; and shows comparison of protocols SEP, LEACH, ESEP, DEEC and ECRSEP  regarding deads nodes, relative to number of rounds. Comparing all these protocols, SEP and LEACH probability based protocols result in approximately equal stability period. As in SEP and LEACH CHs selection is done on probability if LEACH would be considered with homogeneity then there would be a large difference. ESEP with tree levels of heterogeneity and probability based protocol obviously shows better results than SEP and LEACH. Due to availability of more nodes with extra energy ESEP results in increased stability period than SEP and LEACH. The first node of our proposed protocol die after 5000 rounds, achieves greater stability as compared to all protocols discussed in this paper.
Fig. 2 (a) shows that the network life of LEACH is less as compared to all protocols, as it is very sensitive to heterogeneity. Results shows that ECRSEP achieves maximum network lifetime. All nodes are die after 10000 rounds, so network lifetime increases.
Fig. 2 (b) shows the stabile region of the WSN. In a network of heterogeneous nodes, LEACH goes sooner to unstable operation as it is very sensitive to such type of heterogeneity. SEP extend the stability period by aware of heterogeneity through assigning probabilities of CH election weighted by relative initial energy. Due to extended stability throughput of SEP is higher than LEACH. SEP yields longer stability period due to the extra energy of advanced nodes. ESEP have 3 level of heterogeneity, so it have longer stability period than ESEP. Our proposed protocol outperforms all in terms of network stability.
Fig. 2 (c) shows comparison of these protocols regarding throughput, relative to number of rounds  defined as data sent from CH to base station. The throughput of SEP is greater than LEACH in both stable and unstable region. The throughput of ESEP is grater than SEP because of three level of heterogenity. Our proposed protocol ECRSEP beats all protocols however, result shows that throughput of DEEC is maximum as compared to other protocols.
Fig. 2 (d,e,f) shows results, when $\alpha$ = 2 and $m =0.3$. Results in Fig. 2 (d) shows that ECRSEP is achieved maximum network life time as compared to all protocols. All the nodes die after 23000 rounds in our proposed protocol. The network life of other protocols is not more than 8000 rounds. Our protocol beats all these protocols, discussed in this paper.

Fig. 2 (e) shows that nodes die more slowly in ECRSEP, which means that its stability period is increased. LEACH goes sooner to unstable operation as it is very sensitive to such type of heterogeneity. SEP extend the stability period by aware of heterogeneity through assigning probabilities of CH election weighted by relative initial energy. Due to extended stability throughput of SEP is higher than LEACH. SEP yields longer stability period due to the extra energy of advanced nodes. ESEP have 3 level of heterogeneity, so it have longer stability period than ESEP. Our proposed protocol outperforms all in terms of network stability.

Fig. 2 (f) shows that DEEC outclass all protocols in terms of throughput,relative to number of rounds  defined as data sent from CH to base station. Due to supporting heterogeneity the throughput of SEP, ESEP and ECRSEP is higher than LEACH.

\section{Conclusion}
In WSN nodes are not always homogeneous they might be heterogeneous, which increases network complexity. To increase stability and reduce the energy consumption clustering is key technique in WSNs. In this paper, we proposed  ECRSEP protocol and compare our proposed protocol with other protocols of WSN  such as SEP, ESEP, LEACH and DEEC. We conclude that ECRSEP is most suitable when deal with network lifetime and stability.

\end{document}